# Nuclear reaction rates and energy in stellar plasmas : The effect of highly damped modes

Merav Opher, Luis O. Silva, Dean E. Dauger, Viktor K. Decyk and John M. Dawson

*Physics and Astronomy Department*

*University of California, Los Angeles, CA 90095-1547, U.S.A*

## Abstract

The effects of the highly damped modes in the energy and reaction rates in a plasma are discussed. These modes, with wavenumbers $k \gg k_D$, even being only weakly excited, with less than $k_B T$ per mode, make a significant contribution to the energy and screening in a plasma. When the de Broglie wavelength is much less than the distance of closest approach of thermal electrons, a classical analysis of the plasma can be made. It is assumed, in the classical analysis, with $\hbar \to 0$, that the energy of the fluctuations $\hbar \omega \ll k_B T$. Using the fluctuation- dissipation theorem, the spectra of fluctuations with $\hbar \neq 0$ is appreciably decreased. The decrease is mainly for the highly damped modes at high frequencies ($\sim 0.5 - 3 k_B T$). Reaction rates are enhanced in a plasma due to the screening of the reacting ions. This is taken into account by the Salpeter factor, which assumes slow motion for the ions. The implication of including the highly damped modes (with $\hbar \neq 0$) in the nuclear reaction rates in a plasma is discussed. Finally, the investigations presently done on these effects in particle simulations with the *sheet model* and the *multiparticle quantum simulation code* are described.

# I. INTRODUCTION

Plasma effects can drastically modify important phenomena in astronomy such as the energy density in the early universe and the neutrino flux produced in the interior of the Sun. Plasma physics has been extensively studied in the laboratory. However, in astronomy, research on plasmas has received little attention. The lack of a more consistent investigation of the plasma effects is particularly true for plasma effects in basic processes such as reaction rates and equation of state. As all stars possess very hot and dense plasmas, plasma effects on reaction rates and equations of state can significantly affect stellar evolution calculations. These effects can also influence the early universe, especially at the radiation epoch, when the amount of energy in the plasma is comparable to the amount of energy in the radiation (photons).

We discuss in the present article the effects of quasi modes and fluctuations in plasma on energy and reaction rates. Even non-magnetized plasmas contain an oscillating magnetic and electric fields that produce fluctuations in density for example. The fluctuation-dissipation theorem is a powerful description of the fluctuations in plasma for it naturally includes all the fluctuations, such as the quasi-modes.

What are quasi modes or highly damped modes? This denomination refers to modes that are not the normal modes of the plasma. For example, in the case of a non-magnetized plasma, all modes that are not *plasmons* (in the longitudinal component) nor *photons* (in the transverse component). We would expect that the energy in the fluctuations resides mainly in the normal modes, around the plasma frequency for example (in the case of longitudinal component of a non magnetized plasma) and that the contribution of the quasi-modes would be negligible. We show that the contribution for the Debye-Hückel correlation energy for instance, is appreciable. This is crucial since by not taking the limit of $\hbar \to 0$ for classical plasmas, the fluctuations carried by the highly damped modes are strongly affected. Therefore, by decreasing the energy in the highly damped modes, the energy of the plasma is strongly affected. The fluctuations in plasma can affect also the screening and the enhancement of nuclear reaction rates. We will comment on that and on the role of the highly damped modes.

Presently, we are investigating the effects of fluctuations in particle simulations using two approaches: through a classical description with the *sheet model* [1] and through a quantum description with the *multiparticle quantum code* [2]. The plasmas in the interior of stars are dense enough that quantum effects are important. Many analytical calculations to include quantum effects have been attempted. However, all these calculations suffer from several limitations, such as not including for example, non-linear effects. Although many studies have used classical codes to investigate plasma, currently there are no existing quantum particle codes that can directly simulate basic plasma effects [3]. The only existing code deals with very few particles ($\sim 2$). The *multiparticle quantum code* is a cutting edge code to investigate quantum effects in plasma being able to handle hundreds of quantum particles and analyze a wide spectrum of frequencies and wavenumbers.

The outline of this paper is the following: in the Section II, we describe the electromagnetic fluctuations; in Section III and IV, we comment on the effect of the highly damped modes on the energy of a plasma and in reaction rates. We will concentrate on the recent work of Opher and Opher (1999) [4] that used a semi-classical treatment; in Section V, we describe our present investigation of those effects with the *sheet model* and the *multiparticle quantum code*; and finally, in Section VII, we draw the conclusions commenting on the implication of those effects in astronomy and in the early universe.

## II. ELECTROMAGNETIC FLUCTUATIONS

Plasma presents electromagnetic fluctuations even in thermal equilibrium and in a non-magnetized state. Detailed study of the electromagnetic fluctuations has been performed by Dawson [5], Rostoker *et al.* [6], Sitenko *et al.* [7], and Akhiezer *et al.* [8]. Most of these results are compiled in the books of Sitenko and Akhiezer et al. [9,10]. There has been a long tradition in plasma studies to use the analysis of the fluctuations in a plasma as a powerful tool to investigate intrinsic properties of the plasma like the amount of energy, screening and diffusion. For example, the studies of Ichimaru [11], ONeil and Rostoker [12], Lie and Ichikawa [13] of the correlation energy of a plasma; of Dawson [1] and Lin *et al.* [14] on the energy and screening properties, and the analysis of anomalous electron diffusion using the magnetic fluctuations. The fundamental approach to determine the fluctuation properties of any system in the linear regime is the fluctuation-dissipation theorem.

The level of thermal fluctuations in a plasma is related to the dissipative characteristics and temperature of the plasma. Plasma waves are emitted by particles as they move about the plasma and are absorbed (e.g., Landau damped) by the plasma. The balance between emission and absorption leads to a thermal level of field fluctuations. The fluctuation-dissipation theorem describes this level of fluctuations that includes, besides the *normal modes* of the plasma, modes that have a short lifetime, transient modes. For short lived modes as the damping is faster, the excitation has to be faster so that the emission is equal to the absorption in thermal equilibirium.

The quasi-modes (or heavily damped modes) are modes with high wavenumber $k \gg k_D$. Therefore, we are talking about physical phenomena that exist inside the Debye sphere. For the plasmas treated, the closest distance between the particles $d$ is much less than the Debye sphere, i.e., $k_{max} = 1/d \gg k_D$. For example, for $T = 10^5$ $K$, $n = 10^{13}$ $cm^{-3}$, $k_{max} = 1.1 \times 10^4 k_D$. For these plasmas, the number of particles inside the Debye sphere is large $\sim 10^4$ ($g \sim 10^{-4}$). The particles travelling inside the Debye sphere feel the fluctuating electric fields of the bare particles and are continuously scattered by these fields. As they move, they emit plasma waves. These waves have a very small phase velocity (as $k \gg k_D$), less than the thermal velocity of the other particles and are heavily Landau damped. By suffering Landau damping they transfer energy to other modes. The particles excite new modes which are again short lived and the process is repeated. As noted by Tajima et al. [15], "an individual mode (or field) decays by a certain dissipation, giving up energy to particles or other modes, while particles (or other modes) excite new modes and repeat the process and the amount of fluctuations is related to the dissipation."

The fluctuation-dissipation theorem gives for the spectra of the electromagnetic fluctuations in an isotropic plasma in thermal equilibrium [7],

$$\frac{\langle E^2 \rangle_{\mathbf{k}\omega}}{8\pi} = \frac{\hbar}{e^{\hbar\omega/k_B T} - 1} \frac{Im\ \varepsilon_L}{\mid \varepsilon_L \mid^2} + 2\frac{\hbar}{e^{\hbar\omega/k_B T} - 1} \frac{Im\ \varepsilon_T}{\mid \varepsilon_T - \left(\frac{kc}{\omega}\right)^2 \mid^2}, \quad (1)$$

$$\frac{\langle B^2 \rangle_{\mathbf{k}\omega}}{8\pi} = 2\frac{\hbar}{e^{\hbar\omega/k_B T} - 1} \left(\frac{kc}{\omega}\right)^2 \frac{Im\ \varepsilon_T}{\mid \varepsilon_T - \left(\frac{kc}{\omega}\right)^2 \mid^2} \quad (2)$$

where $\varepsilon_L$ and $\varepsilon_T$ are, respectively, the longitudinal and transverse permittivities of the plasma. The first and second term of Eq. (1) are the longitudinal and transverse electric field fluctuations, respectively. The advantage of using the fluctuation-dissipation theorem is that we are able to estimate the energy in the electromagnetic fluctuations for all frequencies and wavenumbers. This estimation requires the dielectric permittivities obtained from the description of the plasma. The calculation includes not only the energy in the fluctuations that are normal modes of the plasma, such as plasmons and photons, but also the energy in the fluctuations that do not propagate, or are short-lived, i.e., quasi-modes. The expressions given in Eqs. (1)-(2) are a quantum description of the plasma [9]. Works such as Opher and Opher [4] are semi-classical descriptions because even though the expression for the fluctuations was quantic, the dielectric permittivities used were classical. For a relativistic or quantum description, it is necessary to obtain adequate dielectric permittivities. This is true also if we are interested in including non-linear effects in the dielectric permittivities.

Recently, Tajima and collaborators [15] revisited the study of fluctuations. They discussed a very interesting suggestion that the magnetic fluctuations could result in a spontaneous magnetic field in a plasma in thermal equilibrium since the magnetic spectrum in a plasma at low frequencies ($\sim 0$) has an intense peak. The plasma description used was a cold plasma description with the collisions described by binary interactions. A later description by Opher and Opher (1997a) [16] including thermal effects (in addition to collisions), showed that in fact, the magnetic spectrum, is not a blackbody spectrum being distorted at low frequencies. However, the high intensity peak at low frequencies is strongly damped.

Extending this study to the transverse electric field, Opher and Opher (1997b) [17] showed that the electromagnetic spectrum (the sum of the magnetic and the transverse spectrum) is different from a blackbody spectrum in vacuum. It has a plateau at low frequencies, possessing more energy than the blackbody spectrum in vacuum. This shows the power of a tool like Fluctuation-Dissipation Theorem.

## III. ON THE ENERGY AND SPECTRA OF THE FLUCTUATIONS FOR A CLASSICAL PLASMA WITH $\hbar \neq 0$

The classical theory states that the energy of a plasma in thermal equilibrium is given by the ideal gas thermal energy, $3/2\ k_B T$ plus a correction due to the interactions of the particles in a plasma, $E_{corr}$. To examine the question of the energy of a plasma through the fluctuations we need to estimate the energy contained in the fluctuations. Therefore, one must integrate Eqs. (1) and (2) over wavenumber and frequency, with a specific dielectric permittivity. This will give us the energy in the electric field (longitudinal and transverse) and in the magnetic field. The quantity that is interesting, in fact, is not the energy in the fluctuations, but how much the fluctuations contribute to the plasma energy. For that, it is necessary to subtract from $\langle E_L^2 \rangle$ the energy of the particles when they are far apart from each other. This give us what is called the correlation energy, $E_{corr}$, of the particles. The energy of the particles that are separated from each other, at infinity, $E_{inf}$, is given by

$$E_{inf} = \frac{n}{2} \int d\mathbf{k} \frac{4\pi e^2}{k^2} \ . \tag{3}$$

Subtracting $E_{inf}$ from the energy in the fluctuations, $\langle E_L^2 \rangle$ give us the correlation energy,

$$E_{corr} = \langle E_L^2 \rangle - E_{inf} \ . \tag{4}$$

The standard classical description [10,12] uses the limit of $\hbar \to 0$ implying that $\hbar \omega \ll k_B T$. In this limit Eqs. (1) and (2) turn out to be proportional to $(k_B T/\omega)$ and can be integrated in wavenumber in a general way with the aid of the Kramers-Kronig relations without even specifying the particular dielectric permittivity. For instance, the longitudinal component turn out to be,

$$\langle E_L^2 \rangle_k = \frac{nk_BT}{2}\left\{1 - \frac{1}{\varepsilon_l(k,0)}\right\} . \tag{5}$$

In the case of a collisionless electron plasma,

$$E_{corr} = -\frac{ne^2}{\pi}\int dk\frac{1}{1+(k/k_D)^2} = -\left(\frac{1}{12\pi}g\right)\left(\frac{3}{2}nT\right)\frac{2}{\pi}ArcTan\left(\frac{k}{k_D}\right) , \tag{6}$$

where $g$ is the plasma parameter. For $k \gg k_D$, we obtain the standard result for the correlation energy, the Debye-Hückel energy (for higher orders in $g$ see [12]),

$$E_{corr} = -\frac{3}{2}nT\left(\frac{1}{12\pi}g\right) . \tag{7}$$

Opher and Opher (1999) [4] recently obtained the interesting result that the energy in a plasma is orders of magnitude higher than the Debye-Hückel result. For example, for $T = 10^5$ $K$ and $n = 10^{13}$ $cm^{-3}$, $E_{corr}$ obtained was $10^5$ larger than $(g/12\pi)(3/2nT)$. They calculated in the same way decribed above, the correlation energy for classical plasmas *but* assumed a finite $\hbar$. That is, they used the general expressions of the fluctuation-dissipation which is a quantum description for the plasma but the dielectric permittivity was a classical one. Therefore, the description of the particles, electrons and ions was classical, but the fields (photons) were quantized. The plasmas studied were classical in the sense that the de Broglie wavelength was much smaller than the closest approach between particles in plasma. In the purely classical case, when $\hbar \to 0$, the Debye-Hückel theory is recovered. This study showed that, quantum corrections even to classical plasma, can be very important.

In this section we summarize our recent article on reviewing this work [18] trying to explain in a simple way, why this result was obtained and what was exactly the role of the highly damped modes. Some of the questions raised by this result were: if this extra energy was due to the cut-off used in the wavenumber integration of the fluctuations; what was the effect of a non-zero $\hbar$; and which modes were responsible for this extra energy: the highly damped modes, such as modes higher that the peak of the blackbody, where $\hbar\omega > k_BT$. Below we summarize our recent analysis [18]:

(a) The first point was to clarify the role of the highly damped modes in classical plasmas and about the cut-off used in the wave number integration. When integrating the spectra, a cut-off is necessary since the fluid description breaks down at small distances. In fact, we show that the extra energy is *not* due to the cut-off in wavenumber. At first, it might seem confusing given that as works like Tajima *et al.* [15] took it as being the Debye wavenumber. The blackbody spectrum is only obtained at high frequencies if we allow the wavenumber to be much higher than the Debye wavenumber. In the case of Tajima *et al.* [15] they had to impose a blackbody spectrum at high frequencies and fit it with the spectra at low frequencies. Opher and Opher [4] used $k_{max}$ being one over the closest distance between the particles in the plasma and obtained the blackbody naturally. However, by examining Eq. (6) it is clear that the Debye-Hückel theory already assumes that $k \gg k_D$. Therefore, even in the pure classical case with $\hbar \to 0$, the highly damped modes are included in the standard treatment and contribute to the energy. This can be seen in Fig. 1 that shows the longitudinal spectra $\langle E_L^2 \rangle_\omega / \langle E_L^2 \rangle_{\omega=0}$ for the classical case with $\hbar \to 0$ [Eq. (1)] vs. $\omega/\omega_{pe}$ for diferent k's. The parameters are $T = 10^5$ K, and $n = 10^{18}$ $cm^{-3}$ and peak of the blackbody is $232\omega_{pe}$. The full curve is for $k = 0.01 k_D$, the dotted for $0.1 k_D$, the dashed for $10 k_D$ and the long-dash is for $k = 100 k_D$. Since as plasmons occupy the phase space for $\omega \sim \omega_{pe}$ and for $k \ll k_D$, we can see the great contribution of the quasi-modes extending to frequency around the peak of the blackbody. The knee in all the curves is due to the Landau damping at $\omega \sim k v_{te}$. As expected, the knee moves to the right for higher k's. This emphasizes that even for a classical calculation, with $\hbar \to 0$, the modes with frequency different from the plasma frequency, carry a substantial portion of the energy.

(b) What is the effect of not taking the limit of $\hbar \to 0$? In Eq. (1) we can see that the presence of $\hbar$ affects the factor $\hbar/(exp(\hbar\omega/k_B T) - 1)$ (since the permittivity used is classical). This factor in the limit of $\hbar \to 0$ turn out to be $k_B T/\omega$. In Fig. 2 we plotted the spectrum $\langle E_L^2 \rangle_\omega / \langle E_L^2 \rangle_{\omega=0}$ vs. $\omega/\omega_{peak}$, where $\omega_{peak}$ is the frequency peak of the blackbody for $k = k_{max}$, in the cases when with $\hbar \to 0$ (full curve) and with $\hbar \neq 0$ (dashed curve). We can see that even at frequencies less than the peak of the blackbody this factor has a major influence. The spectra with finite $\hbar$ substantially decreased from the one with $\hbar \to 0$. This shows clearly that quantum effects are important even for *classical* plasmas at *classical* frequencies ($\hbar\omega \ll k_B T$).

(c) Why Opher and Opher [4] obtained this extra energy? In Fig. 3 we plotted the longitudinal energy, $\langle E_L^2 \rangle_{\omega_{max}}$, as a function of the maximum frequency of the integration, (already integrated in wavenumber). Both the cases with $\hbar \to 0$ (full curve) and $\hbar \neq 0$ (dashed curve) are plotted. As we increase the maximum frequency of integration, the energy increases as expected, approaching a plateau since we ended up summing all the modes in the plasma that carry the energy. The dotted curve is the energy of the particles separated at infinity, $E_{inf}$ calculated classically [Eq.(3)]. The difference between the dotted curve and the full and dashed curve is the correlation energy (Eq. (4)) and it is as expected, a small value $\propto g$, the Debye-Hückel result. The energy spectrum in the case of $\hbar \to 0$ approaches the dotted curve, $E_{inf}$ while the case with $\hbar \neq 0$ does not approach the dotted curve but a different plateau value. This is why Opher and Opher [4] obtained an energy so different from the standard calculation. The great difference in energy with $\hbar \to 0$ and the case with $\hbar \neq 0$ is concentrated in frequencies from $0.5 - 2.0\omega_{peak}$. Also, this imply that the energy of the fluctuations with finite $\hbar \neq 0$ is in fact reduced. So where is the extra energy coming from? The answer is in the energy of the particles at infinity. We have to clarify that if by not taking the limit of $\hbar \to 0$ to be finite, the energy of the particles at infinity is still given by expression Eq. (3). We expect that when the plasma parameter $g$ (or the density) approaches zero, $\Delta \equiv (\frac{3}{2}nT)E_{corr}$ will tend to zero as there are no interactions. In fact lowering $g$, $\Delta_0$ approaches zero, where $\Delta_0$ is the Debye-Hückel result. However, the same does not occur with $\Delta$, where instead it approaches a constant value $\sim 0.06$ approximately the plateau approached by the dotted curve. The energy of a plasma is given by $E = \frac{3}{2}nT + \frac{3}{2}\Delta = E_{id} + (\langle E^2 \rangle - E_{inf})$, where $E_{id}$ is the energy of an ideal gas, and $E_{inf}$ the energy of the particles at infinity. The fact that $\langle E^2 \rangle$ does not approach $E_{id}$ when $g$ approaches zero means that the energy when the particles are infinitely appart, $E_{inf}$, has to be corrected. As we commented before, the integration in wavenumber diverges for both $\langle E^2 \rangle$ and for $E_{inf}$. The correlation energy is finite because both $\langle E^2 \rangle$ and $E_{inf}$ diverge in the same way. This means that we expect that $\Delta$ will be independent of the choice of $k_{max}$ (for $k \gg k_D$). In fact, varying the cut-off around $k_{max}$ we show that the correlation energy for $\hbar \neq 0$ has a strong dependence on $k_{max}$. For the same range of cut-off, $\Delta_0$ ($\hbar \to 0$) is constant [18]. This shows again that the analysis of quantum effects, even in a classical plasma, is very important but that they also need to be included also in the energy of the particles at infinity.

## IV. ELECTROMAGNETIC FLUCTUATIONS EFFECTS IN REACTION RATES

It is known that nuclear reaction rates in plasma proceed at a higher rate than inferred from laboratory experiments. This is because in a plasma, ions are not naked, but are surrounded by electrons that form a shielding cloud around them. The polarization clouds partially screen their charges resulting into a lower Coulomb barrier between them and an enhanced tunneling probability. The first to discuss this effect was Salpeter [19]. He assumed adiabaticity during the collisions, which means that the electrons and ions rearrange themselves so the polarization cloud follows the colliding particles at all times. The reaction rate, $R$, is then found to be increased by the factor $F = exp(U_0/T)$ [20]. (In the static case in the weak screening limit, the enhancing factor is $exp(Z_1 Z_2 e^2/(T R_D))$, the Salpeter factor.)

The exact effect of the plasma on screening has, in the last years, been a source of a great controversy. The solar neutrino problem has lead to a careful study of plasma effects in reaction rates. The solar neutrino problem is the evidence that measurements of electrons neutrinos fluxes in four different experiments (SAGE (Russian-American gallium experiment) [21], Kamiokande [22], Gallex [23] and Homestake [24]), showed that there is a deficit of neutrinos measured compared to the number predicted by the standard solar model. There have been a number of works [25–31] discussing if the screening is in fact described by the Salpeter factor adressing the possibility of dynamic corrections. Carraro *et al.* [25], for example, argued that the enhancement factor should account for the velocity of the colliding particles. Gruzinov [29] argued that there are no dynamic corrections in statistical equilibrium. More recently, Bahcall *et al.* [32] revisited these works pointing out that the screening is given by the Salpeter factor. The Salpeter result is predicted by the Debye-Hückel theory (with $\hbar \to 0$). Here we comment on the work of Opher and Opher (2000b) [31] that examined the role of highly damped modes and how the enhancement ratio is modified when the limit $\hbar \to 0$ is not taken, in a classical plasma. The electromagnetic fluctuations in the plasma act as a background to the interacting ions. They used the same description as in the last Section, a mixture of a classical description of the dielectric permittivity, but not assuming that $\hbar \to 0$, therefore quantizing the fields. The enhancement factor can be written as a function of the background fluctuating potential,

$$F = 1 + Z_1 Z_2 e^2 \langle \phi^2 \rangle \beta^2 + ... = exp\left(Z_1 Z_2 e^2 \langle \phi^2 \rangle \beta^2\right) , \tag{8}$$

where $\beta = 1/T$ ($k_B = 1$). $\langle \phi^2 \rangle$ is related to the fluctuations of the longitudinal electric field in a plasma: $\langle \phi^2 \rangle_k = \langle E^2 \rangle_k / k^2$. When $\langle \phi^2 \rangle_{\omega,k}$ is integrated for all wavenumbers $k$, but only for frequencies up to $\sim \omega_{pe}$ ($\ll T$), the Salpeter potential is recovered. Through the fluctuation-dissipation theorem they evaluated the screening potential. This guarantees the inclusion of the normal and the highly damped modes. Because the reaction rates are exponentially dependent on the screening potential, any tiny changes in the potential will affect drastically reactions with large values of $Z_1 Z_2$ such as $p - {}^7Be$ and $p - {}^{14}N$. In Fig. 4, the potential $\langle \phi^2 \rangle_{\omega,k} / \langle \phi^2 \rangle_{k,\omega=0}$ vs $\omega/\omega_{pe}$ for the reaction $p - {}^{14}N$, with $T = 10^7$ K and $n = 10^{24}$ cm$^{-3}$ is plotted showing the importance of the highly damped modes. The solid curve is for $k = 100 \ k_D$, the dashed curve for $k = 10 \ k_D$ and the dotted curve for $k = k_D$. We see that for $k = k_D$, the potential $\langle \phi^2 \rangle_{\omega,k}$ has non-negligible values only at low frequencies. However, for $k = 10 \ k_D$ and $k = 100 \ k_D$, the potential $\phi_{k,\omega}$ spreads out to high frequencies. For these high values of $k$, instead of dropping abruptly at $\omega \sim \omega_{pe}$, the curves decrease slowly with frequency. There is thus a substantial contribution for $\omega > \omega_{pe}$. Therefore, for large wavenumbers, (i.e., $k > k_D$) fluctuations make a substantial contribution at frequencies higher than the plasma frequency. Opher and Opher (2000b) [31] find that for the plasmas studied, electron-proton plasmas with temperatures $10^7 - 10^8$ $K$ and densities $10^{23} - 10^{28}$ $cm^{-3}$, the reaction rates $p - p$, $^3He - ^3He$, and $p - ^{14}N$ are increased up to $8 - 13\%$. Even if the exact role of quantum correction has to be more carefully examined, this work shows the importance of quantum corrections even for a classical plasma.

Besides this work, another interesting aspect of the fluctuations that could affect the screening is short time fluctuations. In the next Section we present the result from the *sheet model* (valid in the limit of $\hbar \to 0$), that for long time average reproduced the Debye-Hückel prediction Eq. (7), that lead to the Salpeter screening. However, for short time periods, the potential reaches values much higher than the Debye-Hückel average. In fact, this is the same claim made by Shaviv and Shaviv [27] that only the long time average of the electrostatic potential energy of a test particle in thermal equilibrium is given by Salpeter. However, because the reaction rates are instantaneous phenomena, the ions feel the oscillating potential rather than the long time average potential. All those attempts point out the great importance of a better understanding of the role of thermal fluctuations and highly damped modes.

## V. PRESENT WORK: NUMERICAL ANALYSIS WITH THE SHEET MODEL AND MULTIPARTICLE QUANTUM CODE

Presently with the *multiparticle quantum code* and with the *sheet model* we are investigating the fluctuations in the longitudinal electric field. The electrostatic sheet model is one of the oldest and most versatile one-dimensional models for the investigation of plasmas [1,14]. It was used first to investigate electrostatic effects. Later, it was modified to allow motions on the plane of the sheets, as well as perpendicular to them [14]. The one species model [1] considers a plasma composed of a large number of identical charge sheets embedded in a fixed uniform neutralizing background. The sheets are constrained to be perpendicular to, for example, the $x$-axis, and are allowed to move freely in the $x$ direction. They also are allowed to pass freely through each other. The two species model [14] consists of both positively and negatively charged sheets.

Dean E. Dauger (in collaboration with Viktor K. Decyk and J. M. Dawson) developed a multiparticle quantum particle-in-cell code. This code models many particle quantum systems by combining the semiclassical approximation of Feynman path integrals with a parallel plasma particle-in-cell code. Work done so far on the quantum particle-in-cell shows that the primary computational cost of the method is in tracing the classical trajectories through electromagnetic fields. In the implementation, this is handled by a modified particle pusher and particle manager derived from a plasma particle-in-cell code. Combining these with a wavefunction depositor and a plasma codes field solver forms the primary components of the quantum particle-in-cell code. The wavefunction at $t+\Delta t$ is calculated using the action from the classical path and depositing the virtual classical particle contribution on a grid in real space.

Fig. 6 show the preliminary results with the *multiparticle quantum code*, for the longitudinal spectra $\langle E_L^2 \rangle_\omega / \langle E_L^2 \rangle_{\omega=0}$ normalized vs. $\omega/\omega_{pe}$ with 128 quantum particles (summed up to $10^6$ classic trajectories) and with $k_D = 250 k_{Broglie}$. The full curve is for $k = 0.3\ k_D$, the dotted curve, for $k = 39\ k_D$, the dashed for $k = 59\ k_D$, the long dashed for $k = 79\ k_D$ and the dash-dot for $k = 99\ k_D$. This result agrees with the behavior of the theoretical prediction, displayed in Fig. 1, which shows as $k$ increases, an increase of the intensity of the longitudinal spectra and the presence of the Landau damping. The presence of the non-normal modes is shown at high frequencies for $\omega > \omega_{pe}$. The sheet model presents identical results. Fig. 6 shows for the sheet model, the longitudinal spectra $\langle E_L^2 \rangle_k / nk_B T$ vs $k/k_D$. The Debye-Hückel theory predicts the result given by Eq. (5) that for an electron plasma with $\varepsilon_l(k,0) = (k^2 + k_D^2)/k^2$, gives $\langle E_L^2 \rangle_k / nk_B T = 1/((k/k_D)^2 + 1)$. Both the temporal average and the theoretical prediction are plotted as solid lines, being identical. It can be seen that even though the long time average is the Debye-Hückel theory, the instant value of the longitudinal energy oscillates and is very different from the Debye-Hückel value. This points out that the short-time screening of nuclear reaction rates can be very different from Salpeter (given by the Debye-Hückel theory).

## VI. CONCLUSIONS

We discussed in this article the effect of fluctuations in energy and in reaction rates showing the great contribution of highly damped modes even for the standard classical calculation, the Debye-Hückel theory. Recent works showed that the highly damped modes are responsible for an extra energy in the tranverse component, at low frequencies. It also indicates the importance of quantum effects even for classical plasmas. The spectra of fluctuations is significantly reduced at frequencies from $0.5-2.0\omega_{peak}$. It also pointed out the effect of fluctuations and highly damped modes in reaction rates. The results obtained from the numerical simulations with quantum and classical codes, were presented showing the presence of highly damped modes. The sheet model reproduced on average the Debye-Hückel prediction for the level of longitudinal fluctuations. However, it shows that large deviations can occur for short periods of time. This indicate that the Salpeter enhancement factor can be correct only for long time average. All these studies show the rich and varied phenomena that can be studied with fluctuations in plasma. The study with the multiparticle quantum code will be extended to analyse carefully quantum effects in reaction rates. This code is a cutting edge one since there is no particle code actually existing to simulate quantum effects in plasma. Moreover, all these effects can have major consequences in stellar evolution and in the early universe. In the early universe, a modifications of the equation of state and reaction rates can have several consequences. It will modify the expansion rate and affect the production of light elements at the epoch of Primordial Nucleosynthesis, for example. These modifications will also affect stellar evolution. Stellar models are constructed by solving the basic stellar structure equations. The solution of stellar structure equations requires that the opacity, nuclear reaction rates, and equation of state be specified. Even tiny changes will affect drastically the stellar evolution and the determination of the age of the universe though globular clusters, for example [33].

## ACKNOWLEDGMENTS

M.O.would like to thanks DE-FGO3-93ER54224 grant for support.

# REFERENCES

[1] J. M. Dawson, Physics of Fluids 5, 445 (1962).

[2] D. E. Dauger, V. K. Decyk, J. M. Dawson, Bul.Am.Phys.Soc., 45, 75 (2000).

[3] F. Haas, G. Manfredi, and M. Feix, Phys. Rev. E, 62, 2763 (2000).

[4] M. Opher and R.Opher, Phys. Rev. Lett, 82, 4835 (1999).

[5] J. M. Dawson, Adv. Plasma Phys. 1, 1 (1968).

[6] N. Rostoker, R. Aamodt, and O. Eldridge, Ann. Phys. (N.Y.) 31, 243 (1965).

[7] A. G. Sitenko and A. A. Gurin, JETP 22, 1089 (1966).

[8] A I. Akhiezer, I. A. Akhiezer, and A. A. Gurin, JETP Lett. 14, 462 (1961); A. I. Akhiezer, I. A. A. Akhiezer, and A. G. Sitenko, JETP 14, 462 (1961).

[9] A. G. Sitenko, Electromagnetic Fluctuations in Plasma, (Academic Press, New York, 1967).

[10] A. I. Akhiezer, I.A. Akhiezer, R.V.Polovin, A.G. Sitenko, and K.N.Stepanov, Plasma Electrodynamics (Pergamon Press, Oxford, 1975), Vol. 2.

[11] S. Ichimaru, Phys. Rev. A 2, 494 (1970).

[12] T. ONeil and N. Rostoker, Phys. Fluids 8, 1109 (1965).

[13] T. Lie and Y. Ichikawa, Rev. Mod. Phys. 38, 680 (1966).

[14] A. T. Lin, J. M. Dawson, and H. Okuda, Phys. Rev. Lett., 41, 753 (1978).

[15] T. Tajima, S. Cable, and R. M. Kulsrud, Phys. Fluids B, 4, 2338 (1992); T. Tajima, S. Cable, K. Shibata, and R. M. Kulsrud, Astrophys. J. 390, 309 (1992).

[16] M. Opher and R. Opher, Phys. Rev. D, 56, 3296 (1997).

[17] M. Opher and R. Opher, Phys. Rev. Lett. 79, 2628 (1997).

[18] M.Opher and J. M. Dawson, in preparation (2000).

[19] E. E. Salpeter, Australian J. Phys., 7, 373 (1954).

[20] D. D. Clayton, Principles of Stellar Evolution and Nucleosynthesis (New York: McGraw Hill, 1968).

[21] J. N. Abdurashitov *et al.*, Phys. Rev. Lett., 77, 4708 (1996).

[22] Y. Fukuda *et al.*, Phys. Rev. Lett. 77, 1683 (1996).

[23] P. Anselmann *et al.*, Phys. Lett. B 342, 440 (1995).

[24] R. Davis Jr., Prog. Part. Nucl. Phys. 32, 13 (1994).

[25] C. Carraro, A. Schafer, and S. E. Koonin, Astrophys. J. 331, 565 (1988).

[26] M. Bruggen and D. O. Gough, Astrophys. J. 488, 867 (1997).

[27] G. Shaviv and N. J. Shaviv, Astrophys. J., 529, 1054 (2000).

[28] V. N. Tsytovitch, A & A, 356, L57 (2000).

[29] A. V. Gruzinov, Astrophys. J. 496, 503 (1998).

[30] M. Opher and R. Opher, Astrophys. J. 535, 473 (2000).

[31] M. Opher and R. Opher, Nuclear Reaction Rates in a Plasma: The Effect of Highly Damped Modes, submitted for publication in Phys.Rev.Lett. (2000) (astro-ph/000063262).

[32] J. N. Bahcall, L. S. Brown, A. Gruzinov, R. F. Sawyer, astro-ph/0010055 (2000).

[33] C. Chaboyer, P. Demarque, P. J., Kernan, L. M. Krauss, Science, 271, 957 (1996).

## VII. FIGURES

FIGURES

FIG. 1.

The longitudinal spectra $\langle E_L^2 \rangle_\omega / \langle E_L^2 \rangle_{\omega=0}$ vs. $\omega/\omega_{pe}$ for diferent k's. The parameters are $T = 10^5$ K, and $n = 10^{18}$ $cm^{-3}$ and peak of the blackbody is $232\omega_{pe}$. The full curve is for $k = 0.01 k_D$, the dotted for $0.1 k_D$, the dashed for $10 k_D$ and the long-dash is for $k = 100 k_D$.

FIG. 2.

The spectrum $\langle E_L^2 \rangle_\omega / \langle E_L^2 \rangle_{\omega=0}$ vs. $\omega/\omega_{peak}$ for $k = k_{max}$, in the cases when with $\hbar \to 0$ (full curve) and with $\hbar \neq 0$ (dashed curve).

FIG. 3.

The longitudinal energy, $\langle E_L^2 \rangle_{\omega_{max}}$, as a function of the maximum frequency of the integration, (already integrated in wavenumber). The full curve is for $\hbar \to 0$ and the dashed curve is for $\hbar \neq 0$.

FIG. 4.

The potential $\langle \phi^2 \rangle_{\omega,k} / \langle \phi^2 \rangle_{k,\omega=0}$ vs $\omega/\omega_{pe}$ for the reaction $p - {}^{14}N$, with $T = 10^7$ K and $n = 10^{24}$ $cm^{-3}$. The solid curve is for $k = 100\ k_D$, the dashed curve for $k = 10\ k_D$ and the dotted curve for $k = k_D$.

FIG. 5.

The longitudinal spectra $\langle E_L^2 \rangle_\omega / \langle E_L^2 \rangle_{\omega=0}$ vs. $\omega/\omega_{pe}$. The full curve is for $k = 0.3\ k_D$, the dotted curve, for $k = 39\ k_D$, the dashed for $k = 59\ k_D$, the long dashed for $k = 79\ k_D$ and the dash-dot for $k = 99\ k_D$.

FIG. 6.

The longitudinal spectra $\langle E_L^2 \rangle_k / n k_B T$ vs $k/k_D$